\documentclass{article}

\usepackage{arxiv}

\usepackage[utf8]{inputenc} % allow utf-8 input
\usepackage[T1]{fontenc}    % use 8-bit T1 fonts
\usepackage{hyperref}       % hyperlinks
\usepackage{url}            % simple URL typesetting
\usepackage{booktabs}       % professional-quality tables
\usepackage{amsmath}
\usepackage{amsfonts}       % blackboard math symbols
\usepackage{bm}
\usepackage{amsthm}
\usepackage{amssymb}
\usepackage{multirow}
\usepackage{nicefrac}       % compact symbols for 1/2, etc.
\usepackage{microtype}      % microtypography
\usepackage{lipsum}		% Can be removed after putting your text content
\usepackage{graphicx}
\usepackage[square,numbers]{natbib}
\usepackage{doi}
\usepackage[toc,page]{appendix} 

\newtheorem{thm}{Theorem}

\title{Maximum likelihood estimation of the Weibull distribution with reduced bias}

\author{ \href{https://orcid.org/0000-0003-3017-0871}{\includegraphics[scale=0.06]{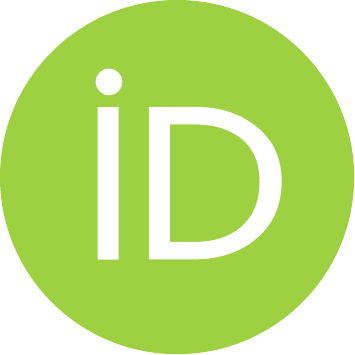}\hspace{1mm}Enes Makalic} \\
	Melbourne School of Population and Global Health\\
	University of Melbourne\\
	Carlton, VIC 3053 \\
	\texttt{emakalic@unimelb.edu.au} \\
	\And
	\href{https://orcid.org/0000-0000-0000-0000}{\includegraphics[scale=0.06]{orcid.pdf}\hspace{1mm}Daniel F.~Schmidt} \\
	Faculty of Information Technology\\
	Monash University\\
	Clayton, VIC 3168 \\
	\texttt{dschmidt@monash.edu} \\
}

\date{}

\renewcommand{\shorttitle}{ML estimation of the Weibull distribution}

\hypersetup{
pdftitle={Maximum likelihood estimation of the Weibull distribution with reduced bias},
pdfauthor={Enes Makalic, Daniel F.~Schmidt},
pdfkeywords={Maximum likelihood, Weibull distribution, bias},
}

\begin{document}
\maketitle

\begin{abstract}
In this short note, we derive a new bias adjusted maximum likelihood estimate for the shape parameter of the Weibull distribution with complete data and type I censored data. The proposed estimate of the shape parameter is significantly less biased and more efficient than the corresponding maximum likelihood estimate, while being simple to compute using existing maximum likelihood software procedures.
\end{abstract}

% keywords can be removed

\section{Introduction}
The Weibull distribution, with probability density function
\begin{equation}
\label{eqn:pdf}
	p(y \vert \bm{\theta}) = \left(\frac{k}{\lambda^k}\right) y^{k-1} \exp\left(-\left(\frac{y}{\lambda}\right)^k\right) ,
\end{equation}
where $\bm{\theta} = (k,\lambda)^\top$ and $k>0$ is the shape parameter and $\lambda>0$ is the scale parameter, is a popular distribution in analysis of survival data. Given data ${\bf y} = (y_1, \ldots, y_n)^\top$, a common approach to estimating the parameters of a Weibull distribution, $\bm{\theta}$, is via the method of maximum likelihood (ML), in which the parameters are set to values that maximise the log-likelihood of the data
\begin{equation}
\label{eqn:complete:nll}
%\ell(\bm{\theta}) = \log p({\bf y} | \bm{\theta}) = -n \log \left(\frac{\lambda^k}{k}\right) + (k-1) \left(\sum_{i=1}^n \log y_i\right) - \sum_{i=1}^n \left(\frac{y_i}{\lambda}\right)^k
\ell(\bm{\theta}) = -n \log \left(\frac{\lambda^k}{k}\right) + (k-1) \left(\sum_{i=1}^n \log y_i\right) - \sum_{i=1}^n \left(\frac{y_i}{\lambda}\right)^k .
\end{equation}
The ML estimate of $\lambda$ is 
\begin{equation}
\label{eqn:ml:k}
%\hat{\lambda}^{k}({\bf y}) = \frac{1}{n} \sum_{i=1}^n y_i^{k}, 
\hat{\lambda}({\bf y}) = \left( \frac{1}{n} \sum_{i=1}^n y_i^{k} \right)^{\frac{1}{k}}, 
\end{equation}
and the ML estimate of $k$, $\hat{k}({\bf y})$, is defined implicitly by the estimating equation
\begin{equation}
\label{eqn:ml:lambda}
 \frac{n}{k} + \sum_{i=1}^n \log y_i - \frac{n \sum_i y_i^k \log y_i}{\sum_i y_i^k} = 0,
\end{equation}
and must be obtained by numerical optimisation. 

The ML estimate of the Weibull distribution scale parameter $\lambda$ has negligible bias, even for relatively small sample sizes. In contrast, the ML estimate of the shape parameter $k$ is known to be strongly biased for small sample sizes. Ross~\cite{Ross94} derived a simple bias-reduction adjustment formula for the ML estimate of $k$
\begin{equation}
\label{eqn:mle:kross}
	\hat{k}_{\rm R}({\bf y}) = \left(\frac{n-2}{n-0.68}\right) \hat{k}_{\rm ML}({\bf y}) ,
\end{equation}
and later extended his approach to censored data~\cite{Ross96}. Hirose~\cite{Hirose99} proposed an alternative bias correction method for data with no censoring that was derived by fitting a non-linear function to simulation results. Teimouri and Nadarajah~\cite{TeimouriNadarajah13} develop improved ML estimates for the Weibull distribution based on record statistics. In contrast, Yang and Xie~\cite{YangXie03} use the modified profile likelihood proposed by Cox and Reid~\cite{CoxReid87,CoxReid92} to derive an alternative ML estimate of $k$ (MLC) from the estimating equation
\begin{equation}
\label{eqn:ml:yangxie}
 \frac{n-2}{k} + \sum_{i=1}^n \log y_i - \frac{n \sum_i y_i^k \log y_i}{\sum_i y_i^k} = 0.
\end{equation}
Using simulations, Yang and Xie showed that their estimate of $k$ is less biased than the ML estimate and more efficient than the estimate (\ref{eqn:mle:kross}) proposed by Ross. In a follow-up paper, Shen and Yang~\cite{ShenYang15} developed a profile ML estimate of $k$ in the case of complete and censored samples, and showed that it outperformed MLC in simulations with complete data.  

In this paper, we introduce new bias adjusted maximum likelihood estimates for the Weibull distribution for both complete and type I censored data. In addition, we derive a novel formula for the Kullback--Leibler (KL)~\cite{KullbackLeibler51} divergence between two Weibull distributions under type I censoring, a result that does not appear to be widely known.
%***DFS: should we add some text here -- like "In this paper, we introduce a new bias correction modification ... ; also do we say we also give a formula for the KL divergence under type I censoring, which does not appear to be widely known***?
%
%
\subsection{Type I Censored Data}
In survival analysis, one typically does not observe complete data and instead has joint realisations of the random variables $(Y = y, \Delta = \delta)$ where $Y = \min (T, c)$ and 
\begin{eqnarray*}
	\Delta &=& {\rm I}(T \leq c) = 
	\begin{cases}
    1, & \text{if } T \leq c \; ({\rm observed\; survival})\\
    0, & \text{if } T > c \; ({\rm observed\; censoring})
	\end{cases},
\end{eqnarray*}
where the random variable $T$ denotes the survival time and $c > 0$ is the fixed censoring time. The data comprises the survival time $T=t$ of an item if this is less than the corresponding censoring time $c$ (i.e., $T \leq c$); otherwise, we only know that the item survived beyond time $c$ (i.e., $T > c$).

The log-likelihood of data $D = \{(y_1, \delta_1), \ldots, (y_n, \delta_n)\}$ under type I censoring is
\begin{align}
\ell(\bm{\theta})
%&= \left(\frac{k}{\lambda^k}\right)^d \exp\left(-\frac{1}{\lambda^k} \sum_{i=1}^n y_i^k \right) \prod_{i=1}^n y_i^{\delta_i (k-1)} \\
&= d \log \left(\frac{k}{\lambda^k}\right) -\frac{1}{\lambda^k} \sum_{i=1}^n y_i^k + \sum_{i=1}^n \log y_i^{\delta_i (k-1)} ,
\label{eqn:censored:nll}
\end{align}
where $d = \sum_{i=1}^n \delta_i$ is the number of uncensored observations. The maximum likelihood (ML) estimate of $\lambda$ is then 
\begin{equation}
\hat{\lambda}^{k}({\bf y}) = \frac{1}{d} \sum_{i=1}^n y_i^{k}, 
\end{equation}
and $\hat{k}({\bf y})$ is obtained from the estimating equation
\begin{equation}
\label{eqn:mle:censored:kscore}
 \frac{d}{k} + \sum_{i=1}^n \delta_i \log y_i - \frac{d \sum_i y_i^k \log y_i}{\sum_i y_i^k} = 0 \, .
\end{equation}

As in the case of complete data, the ML estimate of $k$ for type I censored data has large bias for small sample sizes, and for large amounts of censoring. Based on the modified profile likelihood approach, Yang and Xie~\cite{YangXie03} propose an alternative estimate of $k$ 
\begin{equation}
\label{eqn:ml:yangxie:censored}
 \frac{d-1}{k} + \sum_{i=1}^n \delta_i \log y_i - \frac{d \sum_i y_i^k \log y_i}{\sum_i y_i^k} = 0.
\end{equation}
We note that the above score function requires that $d>1$ to yield a positive estimate for $k$. Yang and Xie demonstrated that their proposed estimate of $k$ is less biased and more efficient than the regular ML estimate. 

Shen and Yang~\cite{ShenYang15} derived a new second- and third-order bias correction formula for the shape parameter of the Weibull distribution without censoring and with general right-censoring models. Although the new estimate is shown to be effective in correcting bias, it must be computed through bootstrap simulation. The same procedure was later extended to include Weibull regression with complete and general right censoring~\cite{ShenYang17}.

More recently, Choi et al~\cite{ChoiEtAl20} examine a different problem of Weibull parameter overestimation caused by mass occurrences of (censored) events in the early time period and develop an expectation maximization (EM) algorithm to reduce bias.

Maximum likelihood estimation of the Weibull distribution under more sophisticated censoring schemes has also been studied. Progressive hybrid censoring and generalized progressively hybrid censored data was examined in \citep{LinEtAl12} and \citep{Zhu20}, respectively. Ng and Wang~\cite{NgWang09} and Teimouri~\cite{Teimouri22} study ML estimation of the Weibull distribution with progressively type I interval censored data. An R package for both progressively type I and type II censored data was developed in~\cite{Teimouri21}. Additionally, ML estimation of the Weibull distribution with generalized type I censored data and block censoring was examined in~\cite{StarlingEtAl021} and \cite{Zhu20b}, respectively.
\section{A simple adjustment to maximum likelihood estimates to reduce estimation bias}
In a landmark paper, Cox and Snell~\cite{CoxSnell68} derived an approximation to the finite sample bias of ML estimates for independent, but not necessarily identically distributed, data (see Appendix~\ref{sec:CoxSnell} for details). The ML estimate with reduced bias, $\bm{\tilde{\theta}}_{\rm ML}$, is given by
\begin{align}
    \bm{\tilde{\theta}}_{\rm ML} 
    &= \hat{\bm{\theta}}_{\rm ML} - \text{Bias}(\hat{\theta}_{\rm ML}) \label{eqn:ml:bias:adjustment},
\end{align}
where the Cox and Snell formula for $\text{Bias}(\hat{\theta}_{\rm ML})$ is given in Appendix~\ref{sec:CoxSnell}, and is evaluated at the usual ML estimate $\hat{\bm{\theta}}_{\rm ML}$. A benefit of this bias approximation formula is that it can be computed even if the ML estimate is not available in closed form. A similar approach to the above was used to derive bias adjusted ML estimates for the unit Weibull distribution~\cite{MenezesEtAl21} and the inverse Weibull distribution~\cite{MazucheliEtAl18} with complete data only. We now extend these results to the Weibull distribution with complete data and Type I censored data.
\begin{thm}
\label{thm:jointpdf}
The finite sample bias of the ML estimate (\ref{eqn:ml:k}) for the Weibull distribution with complete data is 
\begin{align}
\nonumber
\text{\rm Bias}(\hat{k}_{\rm ML}) &= k \left(\frac{18 \left(\pi ^2-2 \zeta (3)\right)}{n \pi ^4}\right) + O(n^{-2})\\
&\approx k \left(\frac{1.3795}{n}\right) \label{eqn:ml:k:adjusted}
%
%\text{\rm Bias}(\hat{\lambda}_{\rm ML}) &= \frac{\lambda  \left(72 (\gamma -1) k \zeta (3)+6 \pi ^2 (5 k+\gamma  (-4 k+\gamma -2)+1)+\pi ^4 (1-2 k)\right)}{2 \pi ^4 k^2 n} + O(n^{-2})\nonumber\\
%
%&\approx \frac{\lambda  (0.5543 - 0.3698 \, k)}{k^2 n} \label{eqn:ml:lambda:adjusted}
\end{align}
where $\zeta(\cdot)$ is the Riemann zeta function. ML estimates of $k$ and $\lambda$ with reduced bias can be obtained from (\ref{eqn:ml:bias:adjustment}). 
\begin{proof}
The proof involves the application of the Cordeiro and Klein~\cite{CordeiroKlein94} approach (see (\ref{eqn:CoxSnell}) and (\ref{eq:Cordeiro:Klein:A}) in Appendix~\ref{sec:CoxSnell}), to the Weibull distribution (\ref{eqn:pdf}). It is well known that expected Fisher information matrix for the Weibull distribution, and its inverse, are given by
\begin{align*}
    {\bf K} &= n \left(
\begin{array}{cc}
 \frac{6 (\gamma -1)^2+\pi ^2 }{6 k^2} & \frac{(\gamma -1) }{\lambda } \\
 \frac{(\gamma -1) }{\lambda } & \frac{k^2 }{\lambda ^2} \\
\end{array}
\right), \\
{\bf K}^{-1} &= \frac{1}{n \pi^2} \left(
\begin{array}{cc}
 6 k^2 & -6 (\gamma -1) \lambda  \\
 -6 (\gamma -1) \lambda  & \frac{\left(6 (\gamma -1)^2+\pi ^2\right) \lambda ^2}{ k^2 } \\
\end{array}
\right),
\end{align*}
where $\gamma \approx 0.5772$ is the Euler--Mascheroni constant. Direct calculation shows that the $2 \times 4$ matrix ${\bf A}$ (see (\ref{eq:Cordeiro:Klein:A}) in Appendix~\ref{sec:CoxSnell}) has entries
\begin{align*}
a_{1,1} &= \frac{n \left(-12 \zeta (3)-3 \gamma  \left(2 \gamma  (\gamma -7)+\pi ^2+16\right)+7 \pi ^2+12\right)}{12 k^3}, \\
a_{1,2} &= a_{2,1} = -\frac{n\left(6 \gamma  (\gamma -4)+\pi ^2+12\right) }{12 k \lambda }, \\
a_{2,2} &= -\frac{n (\gamma  k+k+\gamma -1) }{2 \lambda ^2},\\
a_{1,3} &= -\frac{n \left(6 \gamma  (\gamma -4)+\pi ^2+12\right) }{12 k \lambda }, \\
a_{1,4} &= a_{2,3} = \frac{n (-\gamma  k+3 k+\gamma -1) }{2 \lambda ^2}, \\
a_{2,4} &= -\frac{n (k-1) k^2 }{2 \lambda ^3} .
%\left(
%\begin{array}{cccc}
% \frac{n \left(-12 \zeta (3)-3 \gamma  \left(2 \gamma  (\gamma -7)+\pi ^2+16\right)+7 \pi ^2+12\right)}{12 k^3} & -\frac{\left(6 \gamma  (\gamma -4)+\pi ^2+12\right) n}{12 k \lambda } & -\frac{\left(6 \gamma  (\gamma -4)+\pi ^2+12\right) n}{12 k \lambda } & \frac{(-\gamma  k+3 k+\gamma -1) n}{2 \lambda ^2} \\
% -\frac{\left(6 \gamma  (\gamma -4)+\pi ^2+12\right) n}{12 k \lambda } & -\frac{(\gamma  k+k+\gamma -1) n}{2 \lambda ^2} & \frac{(-\gamma  k+3 k+\gamma -1) n}{2 \lambda ^2} & -\frac{(k-1) k^2 n}{2 \lambda ^3} \\
%\end{array}
%\right)
\end{align*}
Substituting ${\bf K}^{-1}$ and ${\bf A}$ into (\ref{eqn:CoxSnell}) and simplifying completes the proof.
\end{proof}
\end{thm}
From (\ref{eqn:ml:k:adjusted}), we observe that the ML estimate of $k$ is upwardly biased for any finite $n$. A key advantage of the proposed bias adjusted estimate is that it can be trivially computed in any software that implements ML Weibull estimation. We now derive a similar correction for the more complex case of type I censoring.

\begin{figure}[b]
\begin{center}
\includegraphics[width=6.0cm]{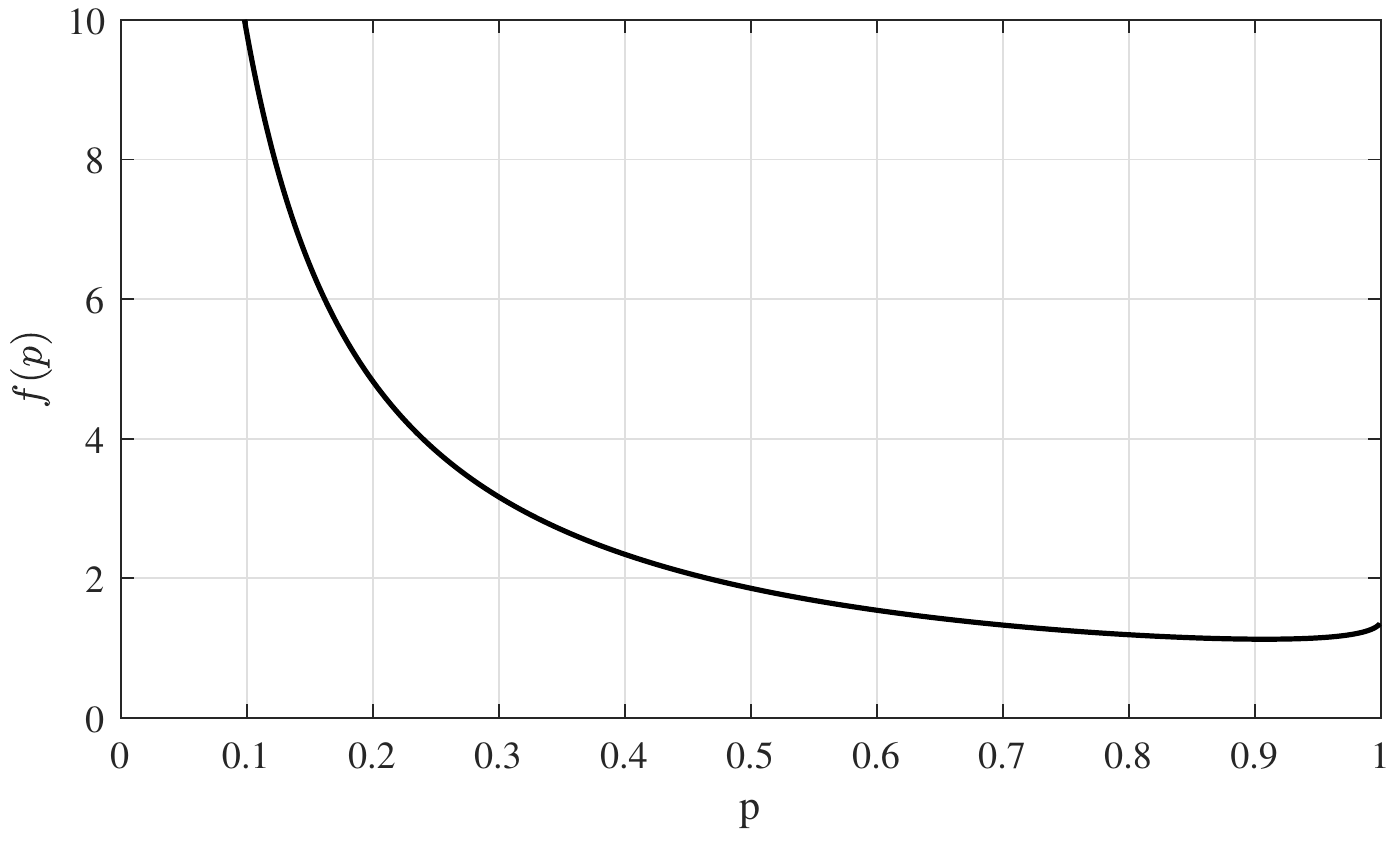}
\end{center}
\caption{Bias adjustment $f(p)$ for the maximum likelihood estimate of the Weibull distribution shape parameter $k$ of as a function of the proportion of uncensored observations $p=1-\exp(-(c/\lambda)^k)$. \label{fig:biasf}}
\end{figure}

%In contrast, the maximum likelihood estimate of $\lambda$ is biased upward for (approximately) $k < 3/2$; otherwise it is biased downward. An advantage of the proposed bias adjusted estimates is that they can be readily computed in any software that implements ML Weibull estimation.
%
%
\begin{thm}
\label{thm:typeI}
The finite sample bias of the maximum likelihood estimate (\ref{eqn:ml:k}) for the Weibull distribution with type I censored data is 
\begin{align}
\text{\rm Bias}(\hat{k}_{\rm ML}) &= k\left(\frac{f(p)}{n}\right) + O(n^{-2}) ,
\end{align}
where $p = 1- \exp(-z_c)$ is the proportion of uncensored observations, $z_c = (c/\lambda)^k$, and
\begin{align}
f(p) &= \frac{-3 \left(2 \gamma_1+\gamma_2\right) \gamma_1 p + \left(6 \gamma_2+\gamma_3\right) p^2+2 \gamma_1^3}{2 \left(\gamma_1^2-\gamma_2 p\right){}^2} ,
\end{align}    
and $\gamma(\cdot,\cdot)$ is the incomplete gamma function
\begin{equation}
\label{eqn:gammainc}
    \gamma(z,x) = \int_0^x t^{z-1} \exp(-t) dt 
\end{equation}
whose $j$-th derivative is
\begin{equation}
\label{eqn:gammainc:deriv}
    \gamma^{(j)} (z, x) = \frac{d^j \gamma(z,x)}{d z^j} .
\end{equation}
For brevity, we use the shorthand notation $\gamma_j \equiv \gamma^{(j)} (1, z_c)$ to denote the $j$-th derivative of the incomplete gamma function evaluated at $(1,z_c)$. As in the case of complete data, the ML estimate of $k$ with reduced bias can be obtained from (\ref{eqn:ml:bias:adjustment}).
\begin{proof}
The expected Fisher information matrix for the Weibull distribution with type I censoring is~\cite{WatkinsJohn04}
\begin{align*}   
{\bf K} &= n \left(
\begin{array}{cc}
 \frac{p+2 \gamma_1+\gamma_2}{k^2} & -\frac{p+\gamma_1}{\lambda } \\
 -\frac{p+\gamma_1}{\lambda } & \frac{k^2 p}{\lambda ^2} \\
\end{array}
\right), \\
{\bf K}^{-1} &= \frac{1}{n(\gamma_2 p - \gamma_1^2)}\left(
\begin{array}{cc}
 k^2 p & \lambda  \left(p+\gamma_1\right) \\
 \lambda  \left(p+\gamma_1\right) & \frac{\lambda ^2 \left(p+2 \gamma_1+\gamma_2\right)}{k^2} \\
\end{array}
\right) .
\end{align*}
By direct calculation we have
\begin{align*}
    a_{1,1} &=  \frac{n \left(2 p+8 \gamma_1+7 \gamma_2+\gamma_3\right)}{2 k^3} , \\
    a_{1,2} &= a_{2,1} = -\frac{n \left(2 p+4 \gamma_1+\gamma_2\right)}{2 k \lambda } , \\
    a_{2,2} &= \frac{n \left(\gamma_1 (k+1)-(k-1) p\right)}{2 \lambda ^2} , \\
    a_{1,3} &= -\frac{n \left(2 p+4 \gamma_1+\gamma_2\right)}{2 k \lambda } , \\
    a_{1,4} &= a_{2,3} = \frac{n \left((3 k-1) p+\gamma_1 (k-1)\right)}{2 \lambda ^2} , \\
    a_{2,4} &= -\frac{n (k-1) k^2 p}{2 \lambda ^3} .
\end{align*}
We note that
\begin{align*}
    \lim_{p \to 1} \gamma_1 = -\gamma, \quad \lim_{p\to 1} \gamma_2 = \gamma ^2+\frac{\pi ^2}{6}, \quad \lim_{p\to 1} \gamma_3 = -\gamma ^3-\frac{\gamma  \pi ^2}{2}+\psi ^{(2)}(1) ,
\end{align*}
where $\psi ^{(2)}(1)$ is the second derivative of the polygamma function evaluated at 1. As expected, the matrix ${\bf A}$ for type I censored data converges to the corresponding matrix with complete data as $p \to 1$. Substituting ${\bf K}^{-1}$ and ${\bf A}$ into (\ref{eqn:CoxSnell}) and simplifying completes the proof. 
\end{proof}
%
%where $f(p)$ is a somewhat lengthy and complicated function of the proportion of uncensored observations, $p = 1-\exp(-\left(c/\lambda )\right)^k)$. A simple approximation to $f(p)$ using rational functions is 
%
%\begin{equation*}
    %f(p) \approx \frac{-580.684 p^3+4690.74 p^2-20743.7 p+18830}{-17026.8 p^2+18804.5 p+1},
%\end{equation*}
%
%with the absolute approximation error being less than $0.003$ for all $0.05 \leq p \leq 0.95$. 
%
%\begin{proof}
%\end{proof}
%
\end{thm}
Figure~\ref{fig:biasf} shows the bias adjustment as a function of the proportion of uncensored observations $p$. As the proportion of uncensored observations $p \to 1$ (i.e., no censoring), $f(p) \to (\approx) 1.3795$ as expected. Additionally, $f(p) \to \infty$ as the proportion of censored data is increased (i.e., $p \to 0$).
\\ \\
{\bf Remark.} As noted in the introduction, the ML estimate of the scale parameter $\lambda$ has negligible bias even for small sample sizes. For complete data, this finite sample bias, computed using the Cox and Snell methodology, is:
\begin{align}
\text{\rm Bias}(\hat{\lambda}_{\rm ML}) 
%&= \frac{\lambda  \left(72 (\gamma -1) k \zeta (3)+6 \pi ^2 (5 k+\gamma  (-4 k+\gamma -2)+1)+\pi ^4 (1-2 k)\right)}{2 \pi ^4 k^2 n} \nonumber \\
%& \quad \quad  + O(n^{-2})\nonumber\\
%
%&\approx \frac{\lambda  (0.5543 - 0.3698 \, k)}{k^2 n} 
%\label{eqn:ml:lambda:adjusted}    \\
&= \lambda \left(\frac{ 1 }{n k^2} \left(\frac{3 (\gamma -1)^2}{\pi ^2}+\frac{1}{2}\right) + \frac{1}{n k} \left(\frac{36 (\gamma -1) \zeta (3)}{\pi ^4}+\frac{15-12 \gamma }{\pi ^2}-1\right) \right)\nonumber \\
& \quad \quad  + O(n^{-2})\nonumber\\
&\approx \lambda \left(\frac{ 0.5543 }{n k^2} - \frac{0.3698}{n k} \right)
\label{eqn:ml:lambda:adjusted}   
\end{align}
where $\gamma \approx 0.5772$ is the Euler--Mascheroni constant. For type I censored data, the finite sample bias is:
\begin{align}
    \text{\rm Bias}(\hat{\lambda}_{\rm ML}) = \lambda \left(\frac{f_1(p)}{n k^2} + \frac{f_2(p)}{n k} \right) + O(n^{-2}) ,
\end{align}
where $p$ is the proportion of uncensored observations, and 
\begin{align*}
f_1(p) &= -\frac{p+2 \gamma_1+\gamma_2}{2 \gamma_1^2-2 \gamma_2 p} ,\\
f_2(p) &= \frac{\left(5 \gamma_2+\gamma_3\right) p^2+\left(-5 \gamma_1^2+\left(\gamma_2+\gamma_3\right) \gamma_1-2 \gamma_2^2\right) p+\left(\gamma_2-2 \gamma_1\right) \gamma_1^2}{2 \left(\gamma_1^2-\gamma_2 p\right){}^2} ,
%
%    f_1(p) &= \frac{70824.7 p^4-148315. p^3+116737 p^2-39692 p-4625.47}{67061.7 p^4-59703.9 p^3-16140.1 p^2-367.234 p+1} \\
%
%    f_2(p) &= \frac{-2448.26 p^4+13146.6 p^3-20164 p^2+11240.7 p-1846.67}{7460.65 p^4-22551.3 p^3+19037 p^2-3748.86 p+1} ,
\end{align*}
%
%with the absolute approximation error for $f_1(\cdot)$ and $f_2(\cdot)$ being less than 0.05 for all $0.05 \leq p \leq 0.95$.
with $\gamma_j \equiv \gamma^{(j)} (1, z_c)$ again denoting the $j$-th derivative of the incomplete gamma function (\ref{eqn:gammainc:deriv}) evaluated at $(1,z_c)$.
\subsection{Simulation}
\label{eqn:sim}
We performed a simulation to examine the finite sample behaviour of the new bias adjusted ML estimates of $k$ for both complete and type I censored data. In all simulations, the scale parameter of the data generating model was set to $\lambda^* = 1$ without loss of generality. Due to the scale invariance of the maximum likelihood estimator and the negligible bias in estimating $\lambda^*$, the simulation results for other values of $\lambda^*$ are expected to yield similar conclusions.

\subsubsection{Complete data}
For each run of the simulation, we generated $n$ data points from the model Weibull$(k^*, \lambda^* = 1)$ where $n = \{10, 20, 50\}$ and the shape parameter was set to $k^* \in \{0.5, 1, 5, 10\}$. Regular maximum likelihood (ML) estimates, our proposed bias adjusted maximum likelihood estimates (MMLE), conditional maximum likelihood estimates (MLC) proposed by Yang and Xie~\cite{YangXie03}, and the profile maximum likelihood estimates of Shen and Yang (MLP)~\cite{ShenYang15} were then computed from the data. We used the second-order bias reduction of Shen and Yang as it was virtually indistinguishable from the third-order formula in our tests. We performed $10^5$ simulations for each combination of $(k^*, n)$ and recorded the average bias, mean squared error and Kullback--Leibler (KL) divergence~\cite{KullbackLeibler51} from the data generating model (see Appendix~\ref{sec:kl}). Simulation results are shown in Table~\ref{tab:results:complete} with the KL results omitted for ease of presentation.

All three bias adjusted ML estimates of $k$ result in a significant reduction in bias compared to the usual ML estimate. Compared to MLC, our proposed estimate yields smaller mean squared error and KL divergence, especially as $k$ increases. The profile ML estimate has a slightly smaller bias than our estimate, while the mean squared error and the KL divergence for the two estimates are virtually identical. Unlike both the MLC and MLP estimates, our bias adjusted ML estimate of $k$ is simple to compute in software via existing Weibull ML estimation procedures and does not require the use of the parametric bootstrap.

\begin{table}[t]
\begin{center}
\begin{tabular*}{\textwidth}{@{\extracolsep{\fill}}ccccccccccc@{\extracolsep{\fill}}}
\toprule
$n$ & $k^*$ & \multicolumn{4}{c}{Bias} & & \multicolumn{4}{c}{Mean Squared Error} \\
  &   & ML & MLC & MLP & MMLE & ~ & ML & MLC & MLP & MMLE \\ 
\cmidrule{1-11}
\multirow{4}{*}{10}  &  0.5 &  0.085 &  0.008 & {\bf  0.001} &  0.004 &  &  0.038 &  0.023 & {\bf  0.022} &  0.023 \\ 
 &  1.0 &  0.170 &  0.017 & {\bf  0.002} &  0.009 &  &  0.151 &  0.092 & {\bf  0.089} &  0.090 \\ 
 &  5.0 &  0.852 &  0.088 & {\bf  0.014} &  0.045 &  &  3.775 &  2.305 & {\bf  2.241} &  2.268 \\ 
 & 10.0 &  1.701 &  0.172 & {\bf  0.026} &  0.087 &  & 15.102 &  9.218 & {\bf  8.978} &  9.079 \\ 
\vspace{-2mm} \\ 
\cmidrule{2-11}
\vspace{-2mm} \\ 
\multirow{4}{*}{20}  &  0.5 &  0.038 &  0.004 & {\bf  0.001} &  0.001 &  &  0.012 &  0.009 & {\bf  0.009} &  0.009 \\ 
 &  1.0 &  0.077 &  0.009 & {\bf  0.001} &  0.003 &  &  0.048 &  0.037 & {\bf  0.037} &  0.037 \\ 
 &  5.0 &  0.382 &  0.042 & {\bf  0.004} &  0.011 &  &  1.203 &  0.928 & {\bf  0.916} &  0.917 \\ 
 & 10.0 &  0.755 &  0.075 & {\bf  0.000} &  0.014 &  &  4.769 &  3.686 & {\bf  3.636} &  3.639 \\ 
\vspace{-2mm} \\ 
\cmidrule{2-11}
\vspace{-2mm} \\ 
\multirow{4}{*}{50}  &  0.5 &  0.014 &  0.002 & {\bf  0.000} &  0.000 &  &  0.004 &  0.003 &  0.003 & {\bf  0.003} \\ 
 &  1.0 &  0.029 &  0.003 & {\bf -0.000} &  0.000 &  &  0.015 &  0.013 &  0.013 & {\bf  0.013} \\ 
 &  5.0 &  0.143 &  0.016 & {\bf  0.001} &  0.001 &  &  0.367 &  0.329 &  0.328 & {\bf  0.327} \\ 
 & 10.0 &  0.290 &  0.036 & {\bf  0.005} &  0.006 &  &  1.458 &  1.308 &  1.300 & {\bf  1.299} \\ 
\vspace{-3mm} \\ 
\bottomrule
\vspace{+1mm}
\end{tabular*}
\caption{Bias and mean squared error for maximum likelihood (ML), conditional maximum likelihood of Yang and Xie (MLC), profile maximum likelihood of Shen and Yang (MLP) and our bias adjusted maximum likelihood (MMLE) estimates of $k^*$ computed over $10^5$ simulations with $\lambda^* = 1$. \label{tab:results:complete}}
\end{center}
\end{table}
\subsubsection{Type I censored data}
We also conducted a similar experiment in the setting of type I censored data. For each iteration of the simulation, we generated $n$ data points from the model Weibull$(k^*, \lambda^* = 1)$ where $n = \{10, 20, 30\}$; the shape parameter was again set to $k^* \in \{0.5, 1, 5, 10\}$. The proportion of uncensored observations was $p \in \{0.3, 0.5, 0.7, 0.9\}$. In addition to the bias and the mean squared error in estimating the shape parameter, we computed the Kullback--Leibler (KL) divergence~\cite{KullbackLeibler51} between the data generating model and each estimated model (see Appendix~\ref{sec:kl}).

The newly proposed bias adjustment estimate of $k$ (MMLE) was  compared to the standard ML estimate, the conditional maximum likelihood estimate (MLC) proposed by Yang and Xie~\cite{YangXie03} and the profile maximum likelihood estimate (MLP) of Shen and Yang~\cite{ShenYang15}. The third-order profile ML estimate suffered from issues regarding numerical stability for small $n$ and large amounts of censoring occasionally resulting in a negative estimate of $k^*$; hence all the comparisons were made with the second-order variant. We restricted the experiments to exclude data sets where the number of uncensored observations $d (=\sum_i \delta_i) < 2$, as MLC may result in negative estimates of $k$ for $d < 2$, though we note this does not cause a problem for our proposed MMLE method. The results of these simulations, averaged over $10^5$ runs for each combination of $(n,p,k^*)$, are shown in Table~\ref{tab:results:censored}, with the KL results omitted for ease of presentation. 

We observe that our MMLE estimate of $k$ is more efficient and less biased than the standard ML estimate of $k$ for all tested values of $(n,p,k^*)$. The conditional ML estimate of $k$ is, in general, more biased and has higher mean squared error compared to the MLP and our MMLE estimates. In terms of bias reduction, the profile ML estimate of $k$ is virtually identical to our MMLE for $n \geq 30$. For small sample sizes ($n = 20$) and higher levels of censoring ($p \leq 0.5$), the MMLE estimate appears superior to MLC and MLP in terms of bias, mean squared error and KL divergence. Additionally, in contrast to the profile ML method, our MMLE estimate is easily computed without the need for numerical simulation, and as such can be easily integrated into any software that implements fitting of the Weibull distribution to complete and type I censored data. 
\begin{table}[tbph]
\addtolength{\tabcolsep}{-4pt}
\begin{minipage}{\textwidth}
\begin{center}
%\begin{tabular*}{ccccccccccccccccccccc} 
%\toprule[1pt]
\begin{tabular*}{\textwidth}{@{\extracolsep{\fill}}cccccccccccccccc@{\extracolsep{\fill}}}
\toprule
$n$ & $p$ & $k^*$ & \multicolumn{4}{c}{Bias} & & \multicolumn{4}{c}{Mean Squared Error} \\
    &     &     & ML & MLC & MLP & MMLE & ~ & ML & MLC & MLP & MMLE \\
\toprule
\multirow{16}{*}{10} & \multirow{4}{*}{ 0.3} &  0.5 &  0.115 &  0.021 &  0.004 & {\bf  0.002} &  &  0.220 &  0.150 &  0.090 & {\bf  0.090} \\ 
 & &  1.0 &  0.228 &  0.040 &  0.005 & {\bf -0.001} &  &  0.605 &  0.401 &  0.303 & {\bf  0.301} \\ 
 & &  5.0 &  1.156 &  0.214 &  0.033 & {\bf  0.008} &  & 14.757 &  9.683 &  7.292 & {\bf  7.248} \\ 
 & & 10.0 &  2.251 &  0.374 &  0.058 & {\bf -0.007} &  & 55.591 & 36.196 & 32.112 & {\bf 28.799} \\ 
 & \multirow{4}{*}{ 0.5} &  0.5 &  0.051 & {\bf  0.001} &  0.001 & -0.003 &  &  0.037 &  0.028 &  0.028 & {\bf  0.028} \\ 
 & &  1.0 &  0.108 &  0.007 &  0.008 & {\bf  0.000} &  &  0.144 &  0.109 &  0.110 & {\bf  0.108} \\ 
 & &  5.0 &  0.556 &  0.051 &  0.053 & {\bf  0.014} &  &  3.672 &  2.785 &  2.785 & {\bf  2.738} \\ 
 & & 10.0 &  1.095 &  0.084 &  0.085 & {\bf  0.009} &  & 15.172 & 11.561 & 11.571 & {\bf 11.377} \\ 
 & \multirow{4}{*}{ 0.7} &  0.5 &  0.034 & {\bf -0.002} &  0.003 & -0.003 &  &  0.019 & {\bf  0.015} &  0.016 &  0.016 \\ 
 & &  1.0 &  0.075 &  0.003 &  0.013 & {\bf  0.002} &  &  0.081 & {\bf  0.065} &  0.068 &  0.066 \\ 
 & &  5.0 &  0.381 &  0.023 &  0.075 & {\bf  0.017} &  &  2.048 & {\bf  1.660} &  1.730 &  1.676 \\ 
 & & 10.0 &  0.681 & {\bf -0.028} &  0.073 & -0.042 &  &  7.474 & {\bf  6.122} &  6.363 &  6.181 \\ 
 & \multirow{4}{*}{ 0.9} &  0.5 &  0.032 & {\bf  0.001} &  0.012 &  0.001 &  &  0.013 &  0.011 &  0.012 & {\bf  0.011} \\ 
 & &  1.0 &  0.063 & {\bf  0.001} &  0.023 &  0.002 &  &  0.051 &  0.042 &  0.045 & {\bf  0.042} \\ 
 & &  5.0 &  0.314 & {\bf  0.001} &  0.111 &  0.008 &  &  1.320 & {\bf  1.082} &  1.165 &  1.083 \\ 
 & & 10.0 &  0.632 & {\bf  0.006} &  0.226 &  0.019 &  &  5.306 & {\bf  4.346} &  4.680 &  4.349 \\ 
\vspace{-2mm} \\ 
\cmidrule{2-12}
\vspace{-2mm} \\ 
\multirow{16}{*}{20} & \multirow{4}{*}{ 0.3} &  0.5 &  0.114 &  0.020 &  0.003 & {\bf -0.000} &  &  0.262 &  0.177 &  0.185 & {\bf  0.102} \\ 
 & &  1.0 &  0.231 &  0.042 &  0.006 & {\bf  0.001} &  &  0.756 &  0.504 &  0.339 & {\bf  0.337} \\ 
 & &  5.0 &  1.149 &  0.207 &  0.036 & {\bf  0.002} &  & 16.835 & 11.123 &  9.821 & {\bf  7.900} \\ 
 & & 10.0 &  2.312 &  0.427 &  0.108 & {\bf  0.014} &  & 82.069 & 54.570 & 59.320 & {\bf 37.053} \\ 
 & \multirow{4}{*}{ 0.5} &  0.5 &  0.054 &  0.003 &  0.003 & {\bf -0.000} &  &  0.037 &  0.028 &  0.028 & {\bf  0.028} \\ 
 & &  1.0 &  0.109 &  0.008 &  0.008 & {\bf  0.000} &  &  0.151 &  0.115 &  0.114 & {\bf  0.112} \\ 
 & &  5.0 &  0.534 &  0.030 &  0.029 & {\bf -0.009} &  &  3.720 &  2.843 &  2.826 & {\bf  2.780} \\ 
 & & 10.0 &  1.091 &  0.081 &  0.080 & {\bf  0.004} &  & 15.178 & 11.578 & 11.507 & {\bf 11.321} \\ 
 & \multirow{4}{*}{ 0.7} &  0.5 &  0.036 &  0.001 &  0.006 & {\bf  0.000} &  &  0.020 & {\bf  0.016} &  0.017 &  0.016 \\ 
 & &  1.0 &  0.072 &  0.001 &  0.011 & {\bf -0.000} &  &  0.078 & {\bf  0.064} &  0.067 &  0.065 \\ 
 & &  5.0 &  0.366 &  0.009 &  0.061 & {\bf  0.003} &  &  1.990 & {\bf  1.619} &  1.692 &  1.640 \\ 
 & & 10.0 &  0.724 &  0.012 &  0.114 & {\bf -0.002} &  &  7.781 & {\bf  6.334} &  6.608 &  6.408 \\ 
 & \multirow{4}{*}{ 0.9} &  0.5 &  0.031 & {\bf  0.000} &  0.011 &  0.001 &  &  0.013 &  0.011 &  0.012 & {\bf  0.011} \\ 
 & &  1.0 &  0.063 & {\bf  0.001} &  0.023 &  0.002 &  &  0.053 & {\bf  0.043} &  0.047 &  0.043 \\ 
 & &  5.0 &  0.309 & -0.004 &  0.106 & {\bf  0.003} &  &  1.298 & {\bf  1.066} &  1.146 &  1.066 \\ 
 & & 10.0 &  0.632 & {\bf  0.006} &  0.225 &  0.019 &  &  5.283 &  4.332 &  4.662 & {\bf  4.332} \\ 
\vspace{-2mm} \\ 
\cmidrule{2-12}
\vspace{-2mm} \\ 
\multirow{16}{*}{30} & \multirow{4}{*}{ 0.3} &  0.5 &  0.065 &  0.007 & {\bf  0.001} & -0.001 &  &  0.056 &  0.041 &  0.037 & {\bf  0.037} \\ 
 & &  1.0 &  0.133 &  0.016 &  0.004 & {\bf -0.000} &  &  0.336 &  0.257 &  0.262 & {\bf  0.179} \\ 
 & &  5.0 &  0.653 &  0.068 &  0.009 & {\bf -0.009} &  &  5.621 &  4.184 &  3.907 & {\bf  3.788} \\ 
 & & 10.0 &  1.334 &  0.161 &  0.045 & {\bf  0.008} &  & 23.086 & 17.180 & 16.069 & {\bf 15.252} \\ 
 & \multirow{4}{*}{ 0.5} &  0.5 &  0.034 &  0.001 &  0.002 & {\bf -0.000} &  &  0.020 & {\bf  0.017} &  0.017 &  0.017 \\ 
 & &  1.0 &  0.070 &  0.004 &  0.006 & {\bf  0.001} &  &  0.081 &  0.067 &  0.068 & {\bf  0.067} \\ 
 & &  5.0 &  0.341 &  0.015 &  0.025 & {\bf  0.000} &  &  2.019 &  1.680 &  1.698 & {\bf  1.679} \\ 
 & & 10.0 &  0.684 &  0.032 &  0.051 & {\bf  0.002} &  &  8.095 & {\bf  6.732} &  6.815 &  6.740\\ 
 & \multirow{4}{*}{ 0.7} &  0.5 &  0.024 &  0.001 &  0.004 & {\bf  0.000} &  &  0.012 & {\bf  0.010} &  0.010 &  0.010 \\ 
 & &  1.0 &  0.047 &  0.001 &  0.007 & {\bf  0.000} &  &  0.046 & {\bf  0.040} &  0.041 &  0.041 \\ 
 & &  5.0 &  0.235 &  0.004 &  0.038 & {\bf  0.001} &  &  1.170 & {\bf  1.020} &  1.050 &  1.030 \\ 
 & & 10.0 &  0.475 &  0.012 &  0.080 & {\bf  0.006} &  &  4.662 & {\bf  4.058} &  4.179 &  4.096 \\ 
 & \multirow{4}{*}{ 0.9} &  0.5 &  0.020 & -0.001 &  0.006 & {\bf -0.000} &  &  0.008 &  0.007 &  0.007 & {\bf  0.007} \\ 
 & &  1.0 &  0.040 & {\bf -0.000} &  0.014 &  0.000 &  &  0.031 &  0.027 &  0.029 & {\bf  0.027} \\ 
 & &  5.0 &  0.198 & -0.003 &  0.065 & {\bf -0.001} &  &  0.778 &  0.684 &  0.716 & {\bf  0.684} \\ 
 & & 10.0 &  0.402 & {\bf -0.000} &  0.137 &  0.005 &  &  3.133 &  2.750 &  2.881 & {\bf  2.748} \\
\vspace{-3mm} \\ 
%\bottomrule[1pt]
\bottomrule
\vspace{+1mm}
\end{tabular*}
\caption{Bias and mean squared error for maximum likelihood (ML),  conditional maximum likelihood of Yang and Xie (MLC),  profile maximum likelihood of Shen and Yang (MLP) and our bias adjusted maximum likelihood (MMLE) estimates of $k^*$ computed over $10^5$ simulations with $\lambda^* = 1$; $p$ denotes the proportion of uncensored observations. \label{tab:results:censored}}
\end{center}
\end{minipage}
\end{table}
\subsection{Real data}
\label{eqn:real:data}
To illustrate the usefulness of our new bias adjusted maximum likelihood estimates, we consider real data on failure voltages from~\cite{Lawless02} (pp., 240) that is also analysed in~\cite{ShenYang15}. The data consists of failure voltages  of two types of electrical cable insulation (type 1 and type 2) of 20 specimens each, and is shown in Table~\ref{tab:voltages} for completeness.

Assuming that the failure voltages can be modelled adequately by the Weibull distribution, the ML estimates of the shape and scale parameters for type 1 cables are $\hat{k}_{\text{ML}} = 9.38$ and $\hat{\lambda}_{\text{ML}} = 47.78$, respectively, and for type 2 cables, the ML estimates are $\hat{k}_{\text{ML}} = 9.14$ and $\hat{\lambda}_{\text{ML}} = 59.12$. The newly proposed MMLE estimates of the shape parameter for type 1 and type 2 cables are easily  obtained from the corresponding ML estimates using (\ref{eqn:ml:k:adjusted}): 
\begin{align*}
\hat{k}_{\text{MMLE}} = 9.38 - 9.38 \left(\frac{1.3795}{20}\right) =  8.74 \quad (\text{type 1}), \\
\hat{k}_{\text{MMLE}} = 9.14 - 9.14 \left(\frac{1.3795}{20}\right) = 8.51 \quad (\text{type 2}) .
\end{align*}
The estimates of the shape parameter proposed in~\cite{YangXie03} and \cite{ShenYang15} are significantly closer to our bias adjusted estimates than to the original maximum likelihood estimates, which exhibit significant upward bias. As expected, bias adjusted estimates of the scale parameter are all approximately equal to the corresponding maximum likelihood estimates.

\begin{table}[t]
\begin{center}
\begin{tabular*}{\textwidth}{@{\extracolsep{\fill}}ccccccccccc@{\extracolsep{\fill}}}
\toprule
\multirow{2}{*}{Type 1} & 32.0 & 35.4 & 36.2 & 39.8 & 41.2 & 43.3 & 45.5 & 46.0 & 46.2 & 46.4 \\ & 46.5 & 46.8 & 47.3 & 47.3 & 47.6 & 49.2 & 50.4 & 50.9 & 52.4 & 56.3 \\
%
%\vspace{-2mm} \\ 
\cmidrule{2-11}
%\vspace{-2mm} \\ 
%
\multirow{2}{*}{Type 2} & 39.4 & 45.3 & 49.2 & 49.4 & 51.3 & 52.0 & 53.2 & 53.2 & 54.9 & 55.5 \\
& 57.1 & 57.2 & 57.5 & 59.2 & 61.0 & 62.4 & 63.8 & 64.3 & 67.3 & 67.7 \\
\bottomrule
\vspace{+1mm}
\end{tabular*}
\caption{Failure voltages (measured in kV/mm) for two types of electrical cable insulation (type 1 and type 2) of 20 specimens each. \label{tab:voltages}}
\end{center}
\end{table}

As a further example, we consider the criminal recidivism data first published in~\cite{RossiEtAl80}. This data consists of 432 survival times of individuals released from Maryland state prisons in the 1970s and followed up for 52 weeks after release (i.e., all censored observations were censored at 52 weeks). Approximately 75\% of the observations were censored, indicating a relatively high degree of censoring. Assuming that the survival times are Weibull distributed, regular ML estimates of the shape and scale parameter were found to be $\hat{k}_{\text{ML}} = 1.37$ and $\hat{\lambda}_{\text{ML}} = 123.68$, respectively. In contrast, our bias adjusted MMLE estimates were $\hat{k}_{\text{MMLE}} = 1.35$ and $\hat{\lambda}_{\text{MMLE}} = 123.68$. As expected, all three bias adjusted estimates examined in this manuscript were similar to the ML estimates due to the relatively large sample size. 

We then randomly sampled 20 observations from the original data without replacement; these were 9, 27, 35, 43 and 46 weeks, with the remaining 15 observations censored at 52 weeks. The ML estimate of the shape parameter of this subsample was found to be $\hat{k}_{\text{ML}} = 1.72$; note this is substantially higher than the ML estimate of $1.37$ obtained on the full data sample. In contrast, our bias adjusted MMLE estimate from this subsample was $\hat{k}_{\text{MMLE}} = 1.39$, which is very close to the estimate obtained on the full sample. This again demonstrates that the ML estimate is strongly upwards biased, especially in the case of smaller sample sizes and high degrees of censoring.

\section{Discussion}
\label{sec:discussion}
Our proposed MMLE approach to first-order bias correction results in improved performance compared to the standard ML estimate in small to medium sample sizes for both complete and type I censored data. The methodology introduced here can also be extended to more sophisticated censoring plans. As an example, consider progressive type I interval censoring (PTIC)~\cite{Aggarwala01}. Here, we have $n$ items entering a life experiment at time $T_0 = 0$. The items are  monitored at $m > 0$ pre-selected times $T_1 < T_2 < \ldots < T_m$ only, with the experiment scheduled to terminate at the last observation time $T_m$. During each inspection time $T_i$ ($i = 1, \ldots, m$), the number of failures $Y_i$ for the time interval $(T_{i-1}, T_i]$ is recorded and $R_i$ surviving items are removed from the experiment at random. The number of removed items may be pre-specified as a percentage $p_i$ of the remaining surviving items $X_i$; that is, $R_i = \lfloor p_i X_i \rfloor$ where $0 < p_i \leq 1$, $\lfloor z \rfloor$ is the largest integer less than or equal to $z$ and $p_m = 1$ as all surviving items are removed from the experiment at time $t_m$. Thus, PTIC may be summarised by $m$ triplets $\{Y_i, R_i, T_i \}_{i=1}^m$.

To obtain first order bias adjusted ML estimates for the Weibull distribution under PTIC, we require the expected Fisher information matrix as well as the expected third order derivatives of the log-likelihood function. A general expression for the expected Fisher information matrix under PTIC is given by Theorem 3.3~\cite{Teimouri22} while Theorem 3.4~\cite{Teimouri22} derives the expected third order derivatives for an arbitrary log-likelihood under PTIC. These two formulas are easily specialised to Weibull distributed survival times. 

A limitation of the Cox and Snell first order bias adjustment approach in the case of PTIC is that an analytic expression for the bias correction is not easily available and the estimator must instead be implemented in software. This is because the expected Fisher information matrix and the expected third order derivatives are somewhat long and cumbersome due to interval censoring and the summation over $m$ monitoring times. To fit Weibull distributed data under PTIC, we recommended the R package \verb@bccp@~\cite{Teimouri21} which features a numerical implementation of the Cox and Snell bias adjustment approach under progressive type I and type II interval censoring for a wide range of distributions.

%\appendix
\begin{appendices}
\section{Cox and Snell approximation}\label{sec:CoxSnell}
Let $\bm{\theta} \in \mathbb{R}^p$, where $p > 0$ is the number of free parameters, which is $p=2$ in the case of the Weibull model. Cox and Snell showed that the bias for the $s$-th element of the ML estimate $\hat{\theta}_{\rm ML}$ can be written as
\begin{equation}
    \left[ \text{Bias}(\hat{\theta}_{\rm ML}) \right]_{s} = \sum_{i=1}^p \sum_{j=1}^p \sum_{l=1}^p \kappa^{s,i} \kappa^{j,l} \left( \frac{1}{2} \kappa_{ijl} + \kappa_{ij,l}\right) + O(n^{-2})
\end{equation}
for $s = 1,\ldots,p$, where the cumulants are
\begin{align}
    \kappa_{ij} &= \mathbb{E}\left\{ \frac{\partial^2 \ell(\bm{\theta})}{\partial \theta_i \partial \theta_j} \right\}, \quad 
    \kappa_{ijl} = \mathbb{E}\left\{ \frac{\partial^3 \ell(\bm{\theta})}{\partial \theta_i \partial \theta_j \partial \theta_l} \right\}, \\
    \kappa_{ij,l} &= \mathbb{E}\left\{ \frac{\partial^2 \ell(\bm{\theta})}{\partial \theta_i \partial \theta_j} \frac{\partial \ell(\bm{\theta})}{\partial \theta_l}\right\},
\end{align}
for $i,j = 1,\ldots, p$ and $\kappa^{i,j}$ is the $(i,j)$-th entry of the \emph{inverse} of the expected Fisher information matrix ${\bf K} = \{ -\kappa_{ij} \}$. Following Cordeiro and Klein~\cite{CordeiroKlein94}, we can compactly write this in matrix notation as
\begin{equation}
\label{eqn:CoxSnell}
    \text{Bias}(\hat{\theta}_{\rm ML}) = {\bf K}^{-1} {\bf A} \text{vec}({\bf K}^{-1}) + O(n^{-2}),
\end{equation}
%
%where the matrix ${\bf A}$ is the $(p \times p^2)$ matrix given by
%
%\begin{align}
%    {\bf A} &= \left[ {\bf A}^{(1)} | {\bf A}^{(2)} | \cdots | {\bf A}^{(p)} \right], \quad 
%    {\bf A}^{(l)} = \{a_{ij}^{(l)} \}, \quad a_{ij}^{(l)} = \frac{1}{2}\kappa_{ijl} + \kappa_{ij,l},
%\end{align}
%
%for $i,j,l = 1,\ldots, p$. Cordeiro and Klein~\cite{CordeiroKlein94} showed that (\ref{eqn:CoxSnell}) holds for dependent data and can be re-written in a form that is simpler to compute: \DFS{This is confusing as the two equations (9) and (11) are identical)}
%
%\begin{equation}
%\label{eqn:CordeiroKlein}
%    \text{Bias}(\hat{\theta}_{\rm ML}) = {\bf K}^{-1} {\bf A} \text{vec}({\bf K}^{-1}) + O(n^{-2}),
%\end{equation}
%
where the matrix ${\bf A}$ is the $(p \times p^2)$ matrix given by
\begin{align}
    \label{eq:Cordeiro:Klein:A}
    {\bf A} &= \left[ {\bf A}^{(1)} \vert {\bf A}^{(2)} \vert \cdots \vert {\bf A}^{(p)} \right], \quad 
    {\bf A}^{(l)} = \{a_{ij}^{(l)} \} \\
    \quad a_{ij}^{(l)} &= \kappa_{ij}^{(l)} - \frac{1}{2} \kappa_{ijl}, \quad \kappa_{ij}^{(l)} = \frac{\partial \kappa_{ij}}{ \partial \theta_l} 
\end{align}
for $i,j,l = 1,\ldots, p$. 

\section{Kullback--Leibler divergence}
\label{sec:kl}

For the case of complete data, the Kullback--Leibler (KL) divergence between the data generating model Weibull$(k_0, \lambda_0)$ and the approximating model Weibull$(k_1, \lambda_1)$ is 
\begin{align*}
{\rm KL}( k_0, \lambda_0 \mid\mid  k_1, \lambda_1) &= \left(\frac{\lambda_0 }{\lambda _1}\right)^{k_1} \left( \frac{k_1}{k_0} \right) \Gamma \left( \frac{k_1}{k_0} \right) + \left(\frac{k_1}{k_0}-1\right) \gamma \\
&+ \log \left(\frac{k_0}{k_1} \left(\frac{\lambda _1}{\lambda_0 }\right)^{k_1}\right)-1   .
\end{align*}
Assuming type I censoring, the KL divergence between two Weibull models Weibull($k_0, \lambda_0$) and Weibull($k_1, \lambda_1$) is
\begin{align*}
\label{eqn:weibull:kl}
{\rm KL}( k_0, \lambda_0 \mid\mid  k_1, \lambda_1) 
&= \exp(-\left(c/\lambda_0\right)^{k_0}) A_1 + \left(\frac{\lambda_0 }{\lambda _1}\right)^{k_1} A_2 \\
&+ \left(1-\frac{k_1}{k_0}\right) A_3 +\log \left(\frac{k_0}{k_1} \left(\frac{\lambda _1}{\lambda_0 }\right)^{k_1}\right)-1 ,
\end{align*}
where
\begin{align*}
A_1 &= \log \left(\frac{k_1 }{k_0}c^{k_1-k_0} \lambda_0^{k_0} \lambda_1^{-k_1}\right)+\left(\frac{c}{\lambda _1}\right){}^{k_1}+1  ,\\
A_2 &= \Gamma \left(\frac{k_1}{k_0}+1\right)-\Gamma \left(\frac{k_1}{k_0}+1,\left(\frac{c}{\lambda }\right)^k\right)  , \\
A_3 &= \text{Ei}\left(-\left(\frac{c}{\lambda_0 }\right)^{k_0}\right)-\gamma ,
\end{align*}
and $\text{Ei}(\cdot)$ is the exponential integral function
\begin{equation}
	\text{Ei}(z) = -\int_{-z}^\infty \frac{\exp(-t)}{t} \, dt .
\end{equation}
\end{appendices}

\bibliographystyle{abbrvnat}
\bibliography{bibliography}

\end{document}